# Angle-resolved photoemission spectroscopy with 9-eV photon-energy pulses generated in a gas-filled hollow-core photonic crystal fiber

H. Bromberger[1,*], A. Ermolov[2], F. Belli[2], H. Liu[1], F. Calegari[1,3], M. Chavez-Cervantes[1], M. T. Li[4], C. T. Lin[4], A. Abdolvand[2], P. St. J. Russell[2], A. Cavalleri[1,5], J. C. Travers[2], I. Gierz[1]

[1]*Max Planck Institute for the Structure and Dynamics of Matter, Center for Free Electron Laser Science, Luruper Chaussee 149, 22761 Hamburg, Germany*

[2]*Max Planck Institute for the Science of Light, Günther-Scharowsky-Str. 1, 91058 Erlangen, Germany*

[3]*Institute for Photonics and Nanotechnologies, IFN-CNR, P.zza Leonardo da Vinci 32, I-20133 Milano, Italy*

[4]*Max Planck Institute for Solid State Research, Heisenbergstr. 1, 70569 Stuttgart, Germany*

[5]*Department of Physics, University of Oxford, Clarendon Laboratory, Parks Rd. Oxford, OX1 3PU, United Kingdom*

*Hubertus.Bromberger@mpsd.mpg.de

**A recently developed source of ultraviolet radiation, based on optical soliton propagation in a gas-filled hollow-core photonic crystal fiber, is applied here to angle-resolved photoemission spectroscopy (ARPES). Near-infrared femtosecond pulses of only few µJ energy generate vacuum ultraviolet (VUV) radiation between 5.5 and 9 eV inside the gas-filled fiber. These pulses are used to measure the band structure of the topological insulator $Bi_2Se_3$ with a signal to noise ratio comparable to that obtained with high order harmonics from a gas jet. The two-order-of-magnitude gain in efficiency promises time-resolved ARPES measurements at repetition rates of hundreds of kHz or even MHz, with photon energies that cover the first Brillouin zone of most materials.**

The photoelectric effect[1] occurs only if a solid is irradiated with photon energies in excess of its work function, which typically lies between 4.5 and 4.8 eV[2]. By measuring the kinetic energy and emission angle of the generated photoelectrons with angle-resolved photoemission spectroscopy (ARPES), the binding energy, $E_B$, and the in-plane momentum, $k_{||}$, of the electrons inside the crystal can be determined. Thus, the technique yields the spectral function of the solid and has evolved into an extraordinarily powerful tool for condensed matter research[2].

The most commonly used light sources for static ARPES are either monochromatized helium lamps (He Iα line at 21 eV and He IIα line at 41 eV) or synchrotron sources, where linewidths below 1 meV have been possible. The typical energy and angle resolution of state-of-the-art analyzers are ΔE < 10 meV and ΔΘ = 0.1°. In the ARPES measurement both the accessible momentum resolution, $\Delta k_{||}$, and maximum in-plane momentum, $k_{max}$, depend on the photon energy. The momentum resolution is thus higher at low photon energies, as shown in recent experiments with 6 eV photons from a continuous wave laser-based ARPES setup, which yielded momentum resolutions as low as $\Delta k_{||}$ = 0.006 Å$^{-1}$ [3].

Pulsed sources are attractive because they can be used for time-resolved ARPES (tr-ARPES) experiments[4-7]. Tr-ARPES probes are often based on the fourth harmonic of femtosecond Ti:sapphire (TiSa) lasers[8]. This process has a conversion efficiency of a few percent, resulting in excellent signal-to-noise ratios due to the high repetition rate of the laser source. However, absorption in the nonlinear crystal (e.g. in beta barium borate - BBO) limits the maximum photon energy range to ≤ 6 eV and restricts the accessible momentum range in an ARPES experiment to $k_{max} \leq 0.6$ Å$^{-1}$.

Recently, Liu et al. increased the accessible momentum range to $k_{max}$ = 0.8 Å$^{-1}$ by frequency doubling a frequency tripled commercial Nd:YVO$_4$ laser in a potassium beryllium fluoroborate (KBBF) crystal, resulting in vacuum ultraviolet (VUV) photon energies of 7 eV (177 nm)[9]. However, the long pulse duration of 10 ps did not allow for time-resolved studies of electron dynamics.

Higher photon energies can be generated by high harmonic generation (HHG) in noble gases pumped by amplified femtosecond lasers. HHG can yield extreme ultraviolet (XUV) radiation over a broad spectrum, extending up to several hundred eV[10-12]. The low conversion efficiency of this process (typically in the range 10$^{-6}$ to 10$^{-5}$) demands, however, high energy femtosecond pulses, which restricts the repetition rate of the laser system to a few kHz and results in comparatively low signal-to-noise ratios for space-charge limited photoemission. Furthermore, in this energy range the momentum resolution is limited and the need for monochromatization further reduces the photon flux.

It has recently been shown that gas-filled, hollow-core photonic crystal fibers with a kagomé pattern (kagomé-PCF)[13], can generate and guide light in the vacuum ultraviolet[14]. By filling the fiber with different gas species at a suitable pressure, the group velocity dispersion at the near infrared pump wavelength of 800 nm can be adjusted to be anomalous, while maintaining sufficient nonlinearity for the formation of high-order optical solitons, tens of femtoseconds in duration with energies of a few μJ. The dynamics of the resulting propagation can be optimized either for generation of a supercontinuum extending from the VUV to the infrared[14, 15] or for emission of femtosecond duration dispersive wave (DW) pulses, wavelength-tunable between 3 eV (400 nm) and 11 eV (113 nm)[13, 15]. In particular,

DW pulses in the VUV (at around 8 eV) with energies in excess of 50 nJ, representing ~1% of the energy in the pump pulses, have been demonstrated[15].

In this letter we show that these DW pulses have sufficient photon flux, signal-to-noise ratio and stability to fulfill the requirements of tr-ARPES studies, and illustrate this by band structure measurements on the topological insulator $Bi_2Se_3$. Importantly, this generation scheme combines the high efficiency of solid-state nonlinear crystals with the spectral coverage of gas sources, which will make it possible to perform photoemission spectroscopy at photon energies that cover the whole Brillouin zone of most materials with high repetition rates.

Figure 1 shows a sketch of the experimental apparatus. We used a commercial 1 kHz, 30 fs multi-pass TiSa amplifier spectrally centered at 1.55 eV (800 nm), continuously attenuated between 0.5 and 7 μJ with a half-wave plate (HWP) and a glass wedge (GW). The pulse duration and linear chirp were optimized using a grating compressor (not shown). The beam pointing was stabilized with an active steering mirror (ASM) and a quadrant photodiode detector. The spatial mode distribution of the laser beam was filtered by focusing into a 150 μm diameter pinhole, after which it was imaged into the 28 or 34 μm diameter kagomé-PCF. The coupling efficiency was always above 80%. The entrance and exit faces of the fiber were located in two separate cells as in previous experiments[14, 15]. The light passed through a fused silica window at the entrance and a $MgF_2$ window at the exit. Neon or hydrogen were pumped into the exit cell, and the pressure in the fiber could be adjusted up to a maximum of 30 bar. The front gas cell was evacuated down to mbar pressure so as to ensure stable launching of light into the kagomé-PCF core. The generated UV light had a divergence of ~10 mrad at the exit from the fiber, and was focused on to the image point of a monochromator used to separate the UV signal from the IR pump light. The first toroidal mirror inside the monochromator collimated the light, after which it was diffracted and dispersed by a grating. The second toroidal mirror focused the dispersed light on to the exit slit, forming an imaging spectrometer.

To characterize the UV signal we used a calibrated channel electron multiplier (CEM) to measure the photon flux as a function of grating angle (i.e., wavelength). A last toroidal mirror focused the light onto the sample. The emitted photoelectrons were then dispersed according to their kinetic energy and emission angle using a hemispherical analyzer (SPECS, Phoibos 150) and counted on a two-dimensional detector array.

Figure 2 shows a representative selection of output spectra generated in the PCF, recorded with the CEM as a function of wavelength. In Fig. 2a the spectrum, generated in a PCF filled with 26.5 bar Ne, is centered at 8.6 eV (145 nm) with a full-width-at-half-maximum (FWHM) of 600 meV (10 nm). The photon flux of $10^6$ photons per pulse ($10^9$ photons per second) is comparable to typical monochromatized HHG sources[16], with the important difference that HHG requires input pulse energies that are three orders of magnitude higher. Note that the photon flux exiting the fiber is expected to be significantly higher (based on previous measurements, more than $10^{10}$ photons per pulse[15]) but is reduced by transmission through the monochromator and refocusing optics. In Fig. 2b we show that the spectrum can be tuned from 5.5 eV (220 nm) to 9 eV (140 nm) by changing from $H_2$ to Ne and adjusting gas pressure and pulse energy. Note that, if He, Ar and Kr are used in addition, continuous tuning from 11 eV (113 nm) to 3 eV (400 nm) is possible[13-15].

To test the suitability of the PCF source for tr-ARPES studies, we performed a proof-of-principle band structure measurement on the topological insulator $Bi_2Se_3$[17]. The sample consisted of stacks of Se-Bi-Se-Bi-Se quintuple layers than could be easily cleaved *in situ* in ultra-high vacuum. The photon energy used for ejecting the photoelectrons was set to 8.6 eV (145 nm). Figure 3 shows the measured photocurrent as a function of binding energy and in-plane momentum, revealing the typical Dirac-cone-like dispersion of quasiparticles at the surface of $Bi_2Se_3$ (dashed red lines are guides to the eye). The states at higher binding energy come from the bulk valence band of $Bi_2Se_3$. The signal-to-noise ratio of the measurement is comparable to that of other 1 kHz sources such as HHG, indicating that the PCF source fulfills all the requirements of tr-ARPES studies.

To summarize, we have applied the recently demonstrated VUV pulsed light source, based on dispersive-wave emission from solitons in a gas-filled kagomé-PCF[14, 15], to ARPES. This source has high conversion efficiency and can be driven with pulses of only a few µJ energy, making it possible to perform measurements at repetition rates of hundreds of kHz using existing commercial pulsed lasers. The source extends the accessible photon energy range beyond what is possible with nonlinear crystals, while retaining momentum resolution and long photoelectron escape depths characteristic of low photon energies. The source is easily tunable between 5.5 and 9 eV with photon fluxes comparable to existing HHG sources that require three orders of magnitude more input pulse energy. The VUV spectrum can be extended out to 11 eV by changing the filling gas to helium. The spectral bandwidth of 600 meV theoretically provides a 3 fs transform-limited pulse, with potential for tr-ARPES studies at extreme time-scales.

We thank Jörg Harms for assisting with the figures.

**References**

[1] H. Hertz, Ann. Phys. 31, 983 (1887)

[2] Stefan Hüfner: Photoelectron Spectroscopy. Principles and Applications. Springer-Verlag, Berlin–Heidelberg–New York, 3rd. edition (2003)

[3] A. Tamai, W. Meevasana, P. D. C. King, C. W. Nicholson, A. de la Torre, E. Rozbicki, and F. Baumberger, Phys. Rev. B 87, 075113 (2013)

[4] F. Schmitt, P. S. Kirchmann, U. Bovensiepen, R. G. Moore, L. Rettig, M. Krenz, J.-H. Chu, N. Ru, L. Perfetti, D. H. Lu, M. Wolf, I. R. Fisher, Z.-X. Shen, Science 321, 1649 (2008)

[5] J. Graf, C. Jozwiak, C. L. Smallwood, H. Eisaki, R. A. Kaindl, D-H. Lee, A. Lanzara, Nature Physics 7, 805 (2011)

[6] T. Rohwer, S. Hellmann, M. Wiesenmayer, C. Sohrt, A. Stange, B. Slomski, A. Carr, Y. Liu, L. Miaja Avila, M. Kalläne, S. Mathias, L. Kipp, K. Rossnagel, M. Bauer, Nature 471, 490 (2011)

[7] J. C. Petersen, S. Kaiser, N. Dean, A. Simoncig, H. Y. Liu, A. L. Cavalieri, C. Cacho, I. C. E. Turcu, E. Springate, F. Frassetto, L. Poletto, S. S. Dhesi, H. Berger, and A. Cavalleri, Physical Review Letters 107, 177402 (2011)

[8] J. Faure, J. Mauchain, E. Papalazarou, W. Yan, J. Pinon, M. Marsi, and L. Perfetti: Full characterization and optimization of a femtosecond ultraviolet laser source for time and angle-resolved photoemission on solid surfaces, Rev. Sci. Instrum. 83, 043109 (2012)

[9] Guodong Liu, Guiling Wang, Yong Zhu, Hongbo Zhang, Guochun Zhang, Xiaoyang Wang, Yong Zhou, Wentao Zhang, Haiyun Liu, Lin Zhao, Jianqiao Meng, Xiaoli Dong, Chuangtian Chen, Zuyan Xu and X. J. Zhou, Rev. Sci. Instrum. 79, 023105 (2008)

[10] A. McPherson, G. Gibson, H. Jara, U. Johann, T. S. Luk, I. A. McIntyre, K. Boyer, C. K. Rhodes, J. Opt. Soc. Am. B 4, 595 (1987)

[11] Ch. Spielmann, N. H. Burnett, S. Sartania, R. Koppitsch, M. Schnürer, C. Kan, M. Lenzner, P. Wobrauschek, F. Krausz, Science 278, 661 (1997)

[12] A. Rundquist, C. G. Durfee III, Z. Chang, C. Herne, S. Backus, M. M. Murnane, H. C. Kapteyn, Science 280, 1412 (1998)

[13] P. St.J. Russell, P. Hölzer, W. Chang, A. Abdolvand, J. C. Travers, Nat. Photon 8, 278–286 (2014)

[14] F. Belli, A. Abdolvand, W. Chang, J. C. Travers, P. St.J. Russell, Optica 2, 292–300 (2015)

[15] Alexey Ermolov, Ka Fai Mak, Michael H. Frosz, John C. Travers, Philip St.J. Russell, arXiv: 1503.09033 [physics] (2015)

[16] F. Frassetto, C. Cacho, C. A. Froud, I.C. E. Turcu, P. Villoresi, W. A. Bryan, E. Springate, L. Poletto, Optics Express, Vol. 19, Issue 20, pp. 19169-19181 (2011)

[17] Hsieh, D.; Y. Xia, D. Qian, L. Wray, J. H. Dil, F. Meier, J. Osterwalder, L. Patthey, J. G. Checkelsky, N. P. Ong, A. V. Fedorov, H. Lin, A. Bansil, D. Grauer, Y. S. Hor, R. J. Cava, M. Z. Hasan, Nature 460 (7259): 1101–1105 (2009)

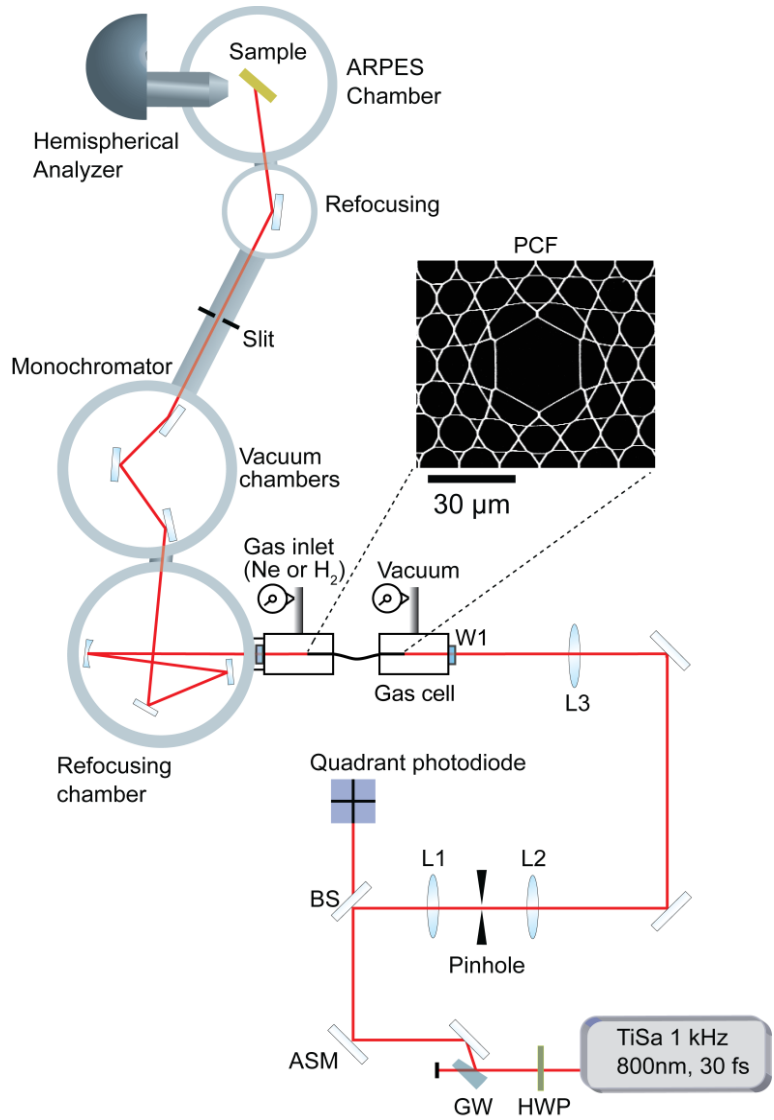

**Figure 1:** Schematic of the setup. Laser light is generated in a TiSa amplifier with 1 kHz repetition rate. A fraction of the light is split off and attenuated to below 10 µJ using a half-wave plate (HWP) and a glass wedge (GW). Beam-pointing stability is achieved using an active steering mirror (ASM) with a quadrant photodiode. The beam profile is cleaned by focusing the beam (with lens L1) onto a pinhole, and then re-collimating it using a second lens (L2). Using a suitable chosen third lens (L3) the beam is then launched into the core (28 µm or 34 µm in diameter) of the kagomé-PCF, which is held between two gas-cells. A positive pressure gradient from vacuum to the set pressure is maintained between the gas-cells. Inside the refocusing chamber the divergent UV light is refocused and steered into the monochromator chamber, where a single grating is used for wavelength selection. Finally, the UV light is focused onto the sample inside the ARPES chamber with a toroidal mirror. Photoelectrons are dispersed with a hemispherical analyzer and counted on a two-dimensional detector.

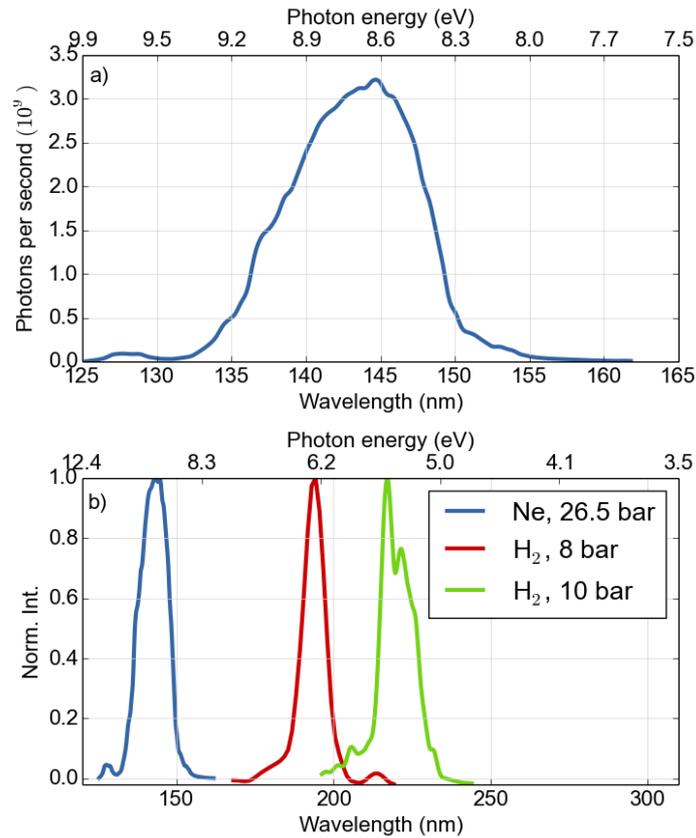

**Figure 2:** UV spectra emitted by gas-filled kagomé-PCF. a) Typical spectrum generated in 26.5 bar Ne, after the monochromator, recorded with a calibrated channel electron multiplier. The count rate is comparable to that usually observed in HHG[16]. b) Normalized spectra at different photon energies, demonstrating the tunability of the source with positive pressure gradient from low vacuum to 26.5 bar of Ne (blue), 8 bar of $H_2$ (red) and 10 bar of $H_2$ (green).

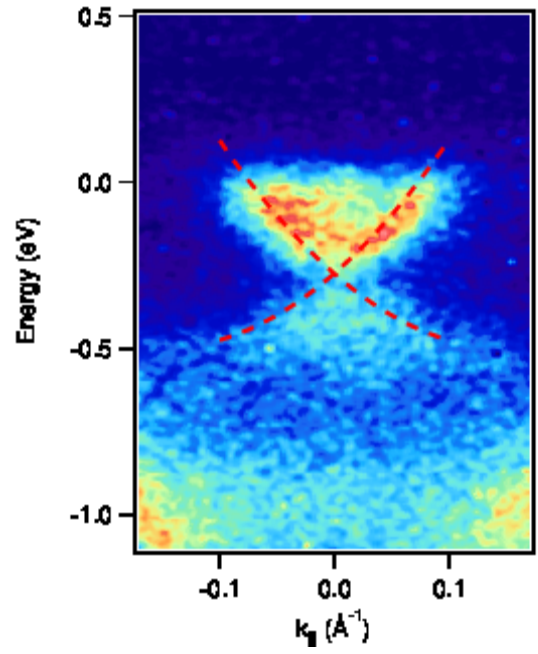

**Figure 3:** ARPES spectrum of Bi$_2$Se$_3$, centered around the Γ-point of the hexagonal Brillouin zone. Dashed red lines following the surface state dispersion are guides to the eye.